\documentclass[prd,twocolumn]{revtex4}
\usepackage{graphicx, epsfig}
\usepackage{color}
\usepackage{mathrsfs}
\usepackage{amsmath}

\newcommand{\be}{\begin{equation}}
\newcommand{\ee}{\end{equation}}
\newcommand{\bea}{\begin{eqnarray}}
\newcommand{\eea}{\end{eqnarray}}

\newcommand{\gapp}{\mathrel{\raise.3ex\hbox{$>$}\mkern-14mu
              \lower0.6ex\hbox{$\sim$}}}
\newcommand{\lapp}{\mathrel{\raise.3ex\hbox{$<$}\mkern-14mu
              \lower0.6ex\hbox{$\sim$}}}

\begin{document}

\title{Hawking Radiation from a Reisner-Nordstr\"om Domain Wall}
\author{Eric Greenwood}
\affiliation{HEPCOS, Department of Physics,
SUNY at Buffalo, Buffalo, NY 14260-1500}
\begin{abstract}

We investigate the effect on the Hawking radiation given off during the time of collapse of a Reisner-Nordstr\"om domain wall. Using the functional Schr\"odinger formalism we are able to probe the time-dependent regime, which is out of the reach of the standard approximations like the Bogolyubov method. We calculate the occupation number of particles for a scalar field and complex scalar field. We demonstrate that the particles from the scalar field are unaffected by the charge of the Reisner-Nordstr\"om domain wall, as is expected since the scalar field doesn't carry any charge, which would couple to the charge of the Reisner-Nordstr\"om domain wall. Here the situation effectively reduces to the uncharged case, a spherically symmetric domain wall. To take the charge into account, we consider the complex scalar field which represents charged particles and anti-particles. Here investigate two different cases, first the non-extremal case and second the extremal case. In the non-extremal case we demonstrate that when the particle (anti-particle) carries charge opposite to that of the domain wall, the occupation number becomes suppressed during late times of the collapse. Therefore the dominate occupation number is when the particle (anti-particle) carries the same charge as the domain wall, as expected due to the Coulomb potential carried by the domain walls. In the extremal case we demonstrate that as time increases the temperature of the radiation decreases until when the domain wall reaches the horizon and the temperature then goes to zero. This is in agreement with the Hawking temperature for charged black holes. 

\end{abstract}

\maketitle

\section{Introduction}

An important question in theoretical physics is that of quantum radiation from collapsing objects. An asymptotic observer, watching gravitational collapse of a massive object, will start registering radiation of quanta coming from the fields excited by the non-trivial metric of the background space-time. As time progresses, the radiation exhibits more and more thermal features, till finally when the horizon is formed the radiation becomes completely thermal. This is in agreement with the fact that the radiation from a pre-existing black hole must be thermal, as originally shown by Hawking, see Ref.\cite{Hawking}.

The typical method for describing the induced radiation of the collapsing object is to use the Bogolyubov method. Here, one considers that the system starts in an asymptotically flat metric, then the system is allowed to evolve to a new asymptotically flat metric. By matching the coefficients between the two asymptotically flat spaces at the beginning and end of the gravitational collapse, the mismatch of these two vacua gives the number of particles produced during the collapse. Hence, what happens in between is beyond the scope of the Bogolyubov method. Therefore the time-evolution of the thermodynamic properties of the collapse cannot be investigated in the context of the Bogolyubov method.

Recently much work has been done with exploring the time-dependent aspects of the induced radiation due to gravitational collapse, see for example Refs.\cite{VachStojKrauss,GreenStoj}. So far, however, these investigations have only included gravitational collapse of massive objects. However, as suggested by the second law of black hole thermodynamics, collapsing objects can contain other observable quantities, such as charge and angular momentum. An important question is, what happens when we introduce these (or a subset of these) additional variables into the mix? Here we investigate the Hawking radiation a charged massive object in the form of a Reisner-Nordstr\"om domain wall. Here the question is, how does the charge effect the time-development of the induced radiation? 

To investigate this question, we shall use the so-called Functional Schr\"odinger formalism (see for example Refs.\cite{VachStojKrauss,GreenStoj,WangGreenStoj}). In general, the Functional Schr\"odinger formalism yields a functional differential equation for the wavefunctional, $\Psi[g_{\mu\nu},X^{\mu},\Phi,{\cal{O}}]$, where $g_{\mu\nu}$ is the metric, $X^{\mu}$ the position of the object, $\Phi$ is a scalar field and ${\cal{O}}$ denotes all the observer's degrees of freedom. Since the Functional Schr\"odinger formalism depends on the observer's degrees of freedom, one can introduce the ``observer" time into the quantum mechanical processes in the form of the Schr\"odinger equation. The wavefunctional is then dependent on the chosen observer's time, hence one can view the quantum mechanical processes of a given system under any choice of space-time foliation. One benefit of this formalism is that one can solve the time-dependent wavefunctional equation exactly using the invariant operator method, in particular  in the present paper. Here we study the propagation of induced radiation, represented as a scalar field and complex scalar field in the background of gravitational collapse, which is governed by the harmonic oscillator equation with a time-dependent frequency. We solve the equations of motion to find the time-dependent wavefunctional for the system. One can then define the occupation number of the induced radiation using the wavefunctional, which is the gaussian overlap between the initial vacuum state and the wavefunction at any given later time. By taking the occupation number to be of the form of the Planck distribution, we can then define $\beta$ which is also time-dependent and fit the temperature of the collapsing object.

In this paper we will investigate the collapse of a charged spherically symmetric infinitely thin domain wall (representing a shell of matter). The details about the collapse will depend on the particular foliation of space-time used to study the system. In this paper we are concerned with the Cosmological view point of the induced radiation during the gravitational collapse, therefore, we adopt the view point of an outside asymptotic observer. Thus, we study the collapse of the domain wall from the view point of a stationary asymptotic observer.

\section{Set Up}

We consider a charged spherical domain wall representing a spherical shell of collapsing matter and charge. The wall is described by only the radial degree of freedom, $R(t)$. The metric is taken to be the solution of Einstein equations for a spherical domain wall with charge. The metric is Schwarzschild outside the wall, as follows from spherical symmetry 
\begin{align}
  ds^2=& -\left(1-\frac{2GM}{r}+\frac{Q^2}{r^2}\right) dt^2\nonumber\\
         &+\left(1-\frac{2GM}{r}+\frac{Q^2}{r^2}\right)^{-1} dr^2 +r^2 d\Omega^2 \ , \ \ r > R(t)
\label{metricexterior}
\end{align}
where $M$ is the mass of the wall, $Q$ is the charge of the wall, and
\begin{equation}
d\Omega^2  = d\theta^2  + \sin^2\theta d\phi^2 \, .
\end{equation}
In the interior of the spherical domain wall, the line element is flat, as expected by Birkhoff's theorem,
\begin{equation}
ds^2= -dT^2 +  dr^2 + r^2 d\theta^2  + r^2 \sin^2\theta d\phi^2  \ ,
\ \ r < R(t)
\label{metricinterior}
\end{equation}
The equation of the wall is $r=R(t)$. The interior time coordinate, $T$, is related to the asymptotic observer time coordinate, $t$, via the proper time of an observer moving with the shell, $\tau$. The relations are
\begin{equation}
\frac{dT}{d\tau} =
      \sqrt{1 + \left (\frac{dR}{d\tau} \right )^2}
      \label{dTdtau}
\end{equation}
and
\begin{equation}
\frac{dt}{d\tau} = \frac{1}{f} \sqrt{f +
         \left ( \frac{dR}{d\tau} \right )^2 }
         \label{dtdtau}
\end{equation}
where
\begin{equation}
f \equiv 1 - \frac{2GM}{R}+\frac{Q^2}{R^2}.
\label{f}
\end{equation}
Upon taking the ratio of Eq.(\ref{dTdtau}) and Eq.(\ref{dtdtau}) one finds 
\be
  \frac{dT}{dt}=\sqrt{f-\frac{(1-f)}{f}R_t^2}
  \label{Tdot}
\ee
where a dot refers to the derivative with respect to the asymptotic observer time $t$.

By integrating the equations of motion for the spherical domain wall, it was shown in Ref.\cite{Lopez} that the mass is a constant of motion and is given by
\be
  M=4\pi \sigma R^2 \left[ \sqrt{1+R_\tau^2} - 2\pi G\sigma R\right] +\frac{Q^2}{2R}.
  \label{Mass}
\ee
where $R_{\tau} = dR/d\tau$, while $\sigma$ is the surface tension (energy density per unit area) of the wall.

\section{Radiation, Semi-Classical Treatment}

Here we consider the radiation given off during gravitational collapse. We will do this for two different cases. First, we will investigate the radiation by coupling a scalar field to the background of the collapsing shell. Second, we will investigate the radiation by coupling a complex (i.e. charged) scalar field to the background of the collapsing shell.

\subsection{Scalar Field}

To investigate the radiation, we consider a scalar field $\Phi$ in the background of the collapsing shell. The scalar field is decomposed into a complete set of basis functions denoted by $\{u_k(r)\}$
\be
  \Phi=\sum_ka_k(t)u_k(r).
  \label{modeEx}
\ee
The exact form of the functions $u_k(r)$ will not be important for us. We will, however, be interested in the wavefunction for the mode coefficients $\{a_k\}$.

The Hamiltonian for the scalar field modes is found by inserting Eq.(\ref{modeEx}), Eq.(\ref{metricexterior}) and Eq.(\ref{metricinterior}) into the action
\be
  S_{\Phi}=\int d^4x\sqrt{-g}\frac{1}{2}g^{\mu\nu}\partial_{\mu}\Phi\partial_{\nu}\Phi.
  \label{rad act}
\ee

Since the metric inside and outside the shell have different forms, we split the action into two parts
\be
  S=S_{in}+S_{out}
\ee
where
\be
  S_{in}=2\pi\int dt\int_0^{R(t)}drr^2\left[-\frac{(\partial_t\Phi)^2}{\dot{T}}+\dot{T}(\partial_r\Phi)^2\right]
  \label{Sin}
\ee
\begin{align}
  S_{out}=2\pi\int dt\int_{R(t)}^{\infty}&drr^2\Big{[}-\frac{(\partial_t\Phi)^2}{1-2GM/r+Q^2/r^2}\nonumber\\
     &\left(1-\frac{2GM}{r}+\frac{Q^2}{r^2}\right)(\partial_r\Phi)^2\Big{]}
     \label{Sout}
\end{align}
where $\dot{T}=dT/dt$ is given in Eq.(\ref{Tdot}). Using the results in Ref.\cite{WangGreenStoj}, one finds that in the non-extremal case (i.e. when the square root in Eq.(\ref{Mass}) is zero) the velocity of the shell, in respect to the asymptotic observer, is given by
\be
  \dot{R}\approx f\sqrt{1-\frac{fR^4}{h^2}}
  \label{Rdot}
\ee
where $h=H/4\pi\mu$, where $H$ is the Hamiltonian of the system (which is a constant of motion), and 
\be
  \mu\equiv1-2\pi\sigma GR_H.
\ee
Using Eq.(\ref{Tdot}) and Eq.(\ref{Rdot}), gives that $\dot{T}$ can be rewritten as
\be
  \dot{T}=f\sqrt{1+(1-f)\frac{R^4}{h^2}}.
  \label{Tdot2}
\ee
If we define the horizons as 
\be
  R_\pm=GM\pm\sqrt{(GM)^2-Q^2}
  \label{Rpm}
\ee
we can then see that as $R\rightarrow R_\pm$, $f\rightarrow0$. From Eq.(\ref{Tdot2}) we see that as $R\rightarrow R_\pm$, $\dot{T}\sim f\rightarrow0$. Therefore the kinetic term in $S_{in}$ diverges as $(R-R_\pm)^{-1}$ in this limit and dominates over the softer logarithmically divergent contribution to the kinetic term from $S_{out}$. Similarly the gradient term in $S_{in}$ vanishes in this limit and is sub-dominant compared to the contribution coming from $S_{out}$. Hence the action becomes
\begin{align}
  S\sim2\pi&\int dt\Big{[}-\frac{1}{f}\int_0^{R_+}drr^2(\partial_t\Phi)^2\nonumber\\
      &+\int_{R_+}^{\infty}drr^2\left(1-\frac{2GM}{r}+\frac{Q^2}{r^2}\right)(\partial_r\Phi)^2\Big{]}
      \label{Approx_act}
\end{align}
where we have changed the limits of integrations to $R_+$ since we are asymptotic observers working in the regime $R(t)\sim R_+$.

Now, we use the expansion in modes in Eq.(\ref{modeEx}) to write
\be
  S=\int dt\left[-\frac{1}{2f}\dot{a}_k{\bf M}_{kk'}\dot{a}_{k'}+\frac{1}{2}a_k{\bf N}_{kk'}a_{k'}\right]
\ee
where ${\bf M}$ and ${\bf N}$ are matrices that are independent of $R(t)$ and are given by
\begin{align}
  {\bf M}_{kk'}=&4\pi\int_0^{R_+}drr^2u_k(r)u_{k'}(r)\\
  {\bf N}_{kk'}=&4\pi\int_{R_+}^{\infty}drr^2\left(1-\frac{2GM}{r}+\frac{Q^2}{r^2}\right)u'_k(r)u'_{k'}(r).
\end{align}

Using the standard quantization procedure, the wavefunction $\psi(a_k,t)$ satisfies
\be
  \left[f\frac{1}{2}\Pi_k({\bf M}^{-1})_{kk'}\Pi_{k'}+\frac{1}{2}a_k{\bf N}_{kk'}a_{k'}\right]\psi=i\frac{\partial\psi}{\partial t}
\ee
where
\be
  \Pi_k=-i\frac{\partial}{\partial a_k}
\ee
is the momentum operator conjugate to $a_k$.

So the problem of radiation from the collapsing domain wall is equivalent to the problem of an infinite set of coupled harmonic oscillators whose masses go to infinity with time. Since the matrices ${\bf M}$ and ${\bf N}$ are symmetric and real (i.e. Hermitian), it is possible to do a principal axis transformation to simultaneously diagonalize ${\bf M}$ and ${\bf N}$, see Ref.\cite{Goldstein}. Then for a single eigenmode, the Schr\"odinger equation takes the form
\be
  \left[-f\frac{1}{2m}\frac{\partial^2}{\partial b^2}+\frac{1}{2}Kb^2\right]\psi(b,t)=i\frac{\partial\psi(b,t)}{\partial t}
  \label{Schrod_t}
\ee
where $m$ and $K$ denote the eigenvalues of ${\bf M}$ and ${\bf N}$, and $b$ is the eigenmode. 

We re-write Eq.(\ref{Schrod_t}) in the standard form
\be
  \left[-\frac{1}{2m}\frac{\partial^2}{\partial b^2}+\frac{m}{2}\omega^2(\eta)b^2\right]\psi(b,\eta)=i\frac{\partial\psi(b,\eta)}{\partial\eta}
  \label{Schrod_eta}
\ee
where
\be
  \eta=\int_0^tdt'f=\int_0^tdt'\left(1-\frac{2GM}{R}+\frac{Q^2}{R^2}\right)
  \label{eta}
\ee
and 
\begin{align}
  \omega^2(\eta)=\frac{K}{mf}=&\frac{K}{m}\frac{1}{1-2GM/R+Q^2/R^2}\nonumber\\
   \equiv&\frac{\omega_0^2}{1-2GM/R+Q^2/R^2}.
  \label{omega}
\end{align}
We have chosen to set $\eta(t=0)=0$. 

To proceed further, we need to choose the background spacetime, i.e. the behavior of $R(t)$. It was shown in Ref.\cite{WangGreenStoj} that the classical solution for the Reisner-Nordstr\"om domain wall in the non-extremal case is given by 
\be
  R(t)\approx R_{+}+(R_0-R_{+})e^{-f_{-}t/R_{+}}
  \label{R_t}
\ee
where $R_+$, which is given in Eq.(\ref{Rpm}), is the outer radius and
\be
  f_{-}=1-\frac{GM-\sqrt{(GM)^2-Q^2}}{R_{+}}.
\ee
Eq.(\ref{R_t}) tells us that $f=1-2GM/R+Q^2/R^2\approx f_{-}e^{-f_{-}t/R_{+}}$ at late times. We are mostly interested in the particle production during this period.  

The spacetime is static at early times and the initial vacuum state for the modes is the simple harmonic oscillator ground state,
\be
  \psi(b,\eta=0)=\left(\frac{m\omega_0}{\pi}\right)^{1/4}e^{-m\omega_0b^2/2}.
\ee
Then the exact solution to Eq.(\ref{Schrod_eta}) at later times is, see Ref.\cite{Pedrosa},
\be
  \psi(b,\eta)=e^{i\alpha(\eta)}\left(\frac{m}{\pi\rho^2}\right)^{1/4}\exp\left[i\frac{m}{2}\left(\frac{\rho_{\eta}}{\rho}+\frac{i}{\rho^2}\right)b^2\right]
  \label{PedWave}
\ee
where $\rho_{\eta}$ denotes the derivative of $\rho(\eta)$ with respect to $\eta$, and $\rho$ is the real solution to the ordinary (though non-linear) auxillary equation
\be
  \rho_{\eta\eta}+\omega^2(\eta)\rho=\frac{1}{\rho^3}
  \label{rho}
\ee
with initial conditions 
\be
  \rho(0)=\frac{1}{\sqrt{\omega_0}}, \hspace{2mm} \rho_{\eta}(0)=0.
  \label{rhoIC}
\ee
The phase $\alpha$ is given by
\be
  \alpha(\eta)=-\frac{1}{2}\int_0^{\eta}\frac{d\eta'}{\rho^2(\eta')}.
  \label{alpha}
\ee

Consider an observer with detectors that are designed to register particles of different frequencies for the free field $\phi$ at early times. Such an observer will interpret the wavefunction of a given mode $b$ at late times in terms of simple harmonic oscillator states, $\{\varphi_n\}$, at the final frequency $\bar{\omega}$. The number of quanta in eigenmode $b$ can be evaluated by decomposing the wavefunction in Eq.(\ref{PedWave}) in terms of the states, $\{\varphi_n\}$, and by evaluating the occupation number of that mode. To implement this evaluation, we start by writing the wavefunction for a given mode at time $t>t_f$ in terms of the simple harmonic oscillator basis at $t=0$. This is given by
\be
  \psi(b,t)=\sum_nc_n(t)\varphi_n(b)
\ee
where
\be
  c_n=\int db\varphi_n^*(b)\psi(b,t)
  \label{c_n}
\ee
which is an overlap of a Gaussian with the simple harmonic oscillator basis functions. The occupation number at eigenfrequency $\bar{\omega}$ (i.e. in the eigenmode $b$) by the time $t>t_f$, is given by the expectation value
\be
  N(t,\bar{\omega})=\sum_nn\left|c_n\right|^2.
\ee
The occupation number in the eigenmode $b$ is then given by (see Appendix \ref{ch:OccNumS})
\be
  N(t,\bar{\omega})=\frac{\bar{\omega}\rho^2}{\sqrt{2}}\left[\left(1-\frac{1}{\bar{\omega}\rho^2}\right)^2+\left(\frac{\rho_{\eta}}{\bar{\omega}\rho}\right)^2\right].
  \label{occupNum}
\ee

Here we will consider only the situation when $(GM)^2>Q^2$, since the case where $Q^2>(GM)^2$ is not as important for the study of Hawking radiation. This is because in the case when $Q^2>(GM)^2$, this represents and oscillating system (due to the repulsive Coulomb potential) and a black hole is never formed, see Ref.\cite{WangGreenStoj}. This occurs since the exponential in Eq.(\ref{R_t}) has an imaginary term, which causes the oscillation of the domain wall. In the case $(GM)^2>Q^2$ we have gravitational collapse and the domain wall will collapse to form a black hole. Here we consider when the charge is large, close to that of $(GM)$. As discussed in Ref.\cite{WangGreenStoj}, when $(GM)^2>>Q^2$, $f$ reduces to $B$, which is given by
\be
  B=1-\frac{2GM}{R}.
\ee
This is the case studied in Ref.\cite{VachStojKrauss} from the view point of the asymptotic observer and in Ref.\cite{GreenStoj} from the view point of the infalling observer (one who is falling in with the shell).   

By calculating $\dot{N}$, it can be checked that $N$ remains constant for $t<0$ and also $t>t_f$. Hence all the particle production occurs for $0\geq t\geq t_f$. There is a possibility that the particle production is due to discontinuities in the derivative of $R$ at $t=0,t_f$. However, as we shall see below, the particle number grows with increasing $t_f$, while the discontinuity at $t=0$ is fixed, and that at $t=t_f$ gets weaker. This indicates that particle production occurs only during $0<t<t_f$ and is a consequence of the gravitational collapse. Now we can take the $t_f\rightarrow\infty$ limit. In Appendix \ref{rho_annal} we have shown that $\rho$ remains finite but $\rho_{\eta}\rightarrow-\infty$ as $t>t_f\rightarrow\infty$, provided $\omega_0\not=0$. However, we are interested in the behavior of $N$ for fixed frequency, $\bar{\omega}$. Since $\bar{\omega}=\omega_0e^{f_{-}t/2R_{+}}$, $t\rightarrow\infty$ also implies $\omega_0\rightarrow0$. From the discussion in Appendix \ref{rho_annal}, we also know that $\rho\rightarrow\infty$ as $\omega_0\rightarrow0$. 

Therefore the occupation number at any frequency diverges in the infinite time limit when backreaction is not taken into account.

\begin{figure}[htbp]
\includegraphics{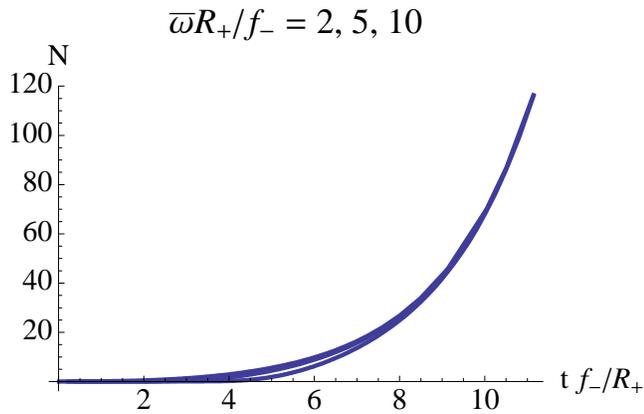}
\caption{Here we plot $N$ versus $tf_-/R_{+}$ for various fixed values of $\bar{\omega}R_{+}/f_-$ for the scalar field case, with $Q^2=0.4^2(GM)^2$. The curves are lower for higher $\bar{\omega}R_{+}/f_-$.}
\label{Nvst_GM}
\end{figure}

In Figure \ref{Nvst_GM} we plot $N$ versus $tf_-/R_{+}$ for various fixed values of $\bar{\omega}R_{+}/f_-$. Here we chose that $Q^2=0.4^2(GM)^2$. We see that the occupation number at any frequency diverges in the infinite time limit when backreaction is not taken into account. 

\begin{figure}[htbp]
\includegraphics{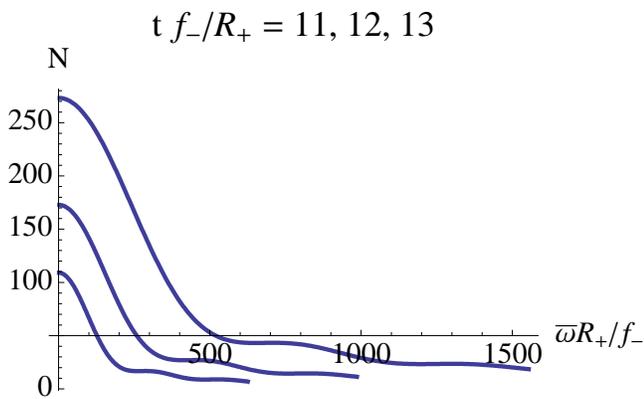}
\caption{Here we plot $N$ versus $\bar{\omega}R_{+}/f_-$ for various fixed values of $tf_-/R_{+}$ for the scalar field case, with $Q^2=0.4^2(GM)^2$. The occupation number at any frequency grows as $tf_-/R_{+}$ increases.}
\label{Nvsw_GM}
\end{figure}

We have also numerically evaluated the spectrum of mode occupation numbers at any finite time and show the results in Figure \ref{Nvsw_GM} for several different values of $tf_-/R_{+}$. We compare the curves in Figure \ref{Nvsw_GM} with the occupation numbers for the Planck distribution
\be
  N_P(\omega)=\frac{1}{e^{\beta\omega}-1}
  \label{N_P}
\ee
where $\beta$ is the inverse temperature. It is clear that the spectrum of occupation numbers is non-thermal. In particular there is no singularity in $N$ at $\omega=0$ at finite time, there are oscillations in $N$. As $tf_-/R_+\rightarrow\infty$, the peak at $\omega=0$ does diverge and the distribution becomes more and more thermal. 

\begin{figure}[htbp]
\includegraphics{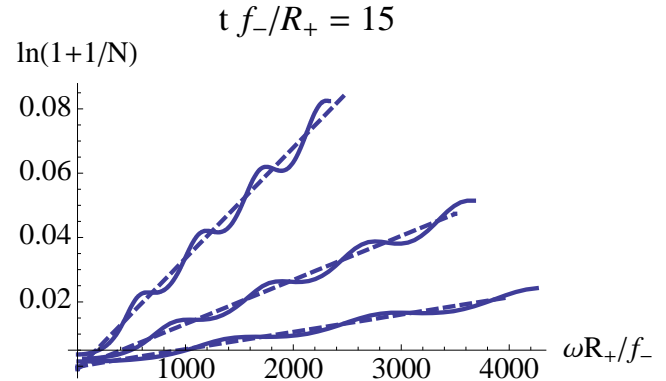}
\caption{Here we plot $\ln(1+1/N)$ versus $\bar{\omega}R_{+}/f_-$ for $tf_-/R_+=15$ for the scalar field case, with $Q^2=0.4^2(GM)^2$. The dashed line shows $\ln(1+1/N_P)$ versus $\bar{\omega}R_{+}/f_-$ where $N_P$ is a Planck distribution. The slope gives $\beta^{-1}$ and the temperature in Eq.(\ref{temp}).}
\label{LnNvsw_GM}
\end{figure}

Now from Eq.(\ref{Schrod_eta}), since the time derivative of the wavefunction on the right-hand side is with respect to $\eta$, $\omega$ is the mode frequency with respect to $\eta$ and not with respect to time $t$. Eq.(\ref{eta}) tells us that the frequency in $t$ is $1-2GM/R+Q^2/R^2$ times the frequency in $\eta$, and at time $t_f$, this implies
\be
  \omega^{(t)}=f_{-}e^{-f_{-}t/R_{+}}\bar{\omega}
  \label{omega_t}
\ee
where the superscript $(t)$ on $\omega$ refers to the fact that this frequency is with respect to $t$. This rescaling of the frequency implies that the temperature for the asymptotic observer (with time coordinate $t$) can be obtained by finding the ``best fit temperature" $\beta^{-1}$ and then rescaling by $1-2GM/R+Q^2/R^2$. So the temperature seen by the asymptotic observer is
\be
  T=f_{-}e^{-f_{-}t/R_{+}}\beta^{-1}(t_f).
  \label{temp}
\ee
(The temperature $T$ is not to be confused with the time coordinate within the spherical domain wall, also denoted by $T$.) We can fit the thermal spectrum to the collapsed spectrum of Figure \ref{Nvsw_GM}, as shown in Figure \ref{LnNvsw_GM} to obtain the temperature of radiation. Here we see that the inverse temperature decreases as $tf_-/R_+$ increase, meaning that the temperature increases over time. 

We thus see that in the context of the Schr\"odinger formalism there is evidence of Hawking-like, but non-thermal radiation emitted during gravitational collapse for the Reisner-Nordstr\"om domain wall before any event horizon is formed. 

Here we make some general comments on the results of this section. Comparing the results from this section with those found in Ref.\cite{VachStojKrauss}, we see that the presence of the charge has qualitatively no effect when considering the radiation in the form of a neutral scalar particle. This is expected since the scalar field is only influenced by the gravity, not by the charge, since the scalar field itself has no charge. Therefore one would expect that there is no effect from the charge. There are, of course, minor quantitative difference due to differences in the background metric (Schwarzschild versus Reisner-Nordstr\"om), such as the surface gravity. However, for this analysis they are not important. If one wants to investigate the effect of the charge, one would need to consider a complex scalar field. This is done in the next section.

\subsection{Complex Scalar Field}

The above analysis is consistent with what one would expect when coupling a scalar field to the background of the Reisner-Nordstr\"om domain wall, since the scalar field has no charge. Therefore, it is advantageous to investigate the effect of adding charge to the scalar field and investigating the effect of the charge. This is done by considering a complex scalar field coupled to the background of the Reisner-Nordstr\"om domain wall. Thus the action in Eq.(\ref{rad act}) becomes
\be
  S_{\Phi}=\int d^4x\sqrt{-g}\frac{1}{2}g^{\mu\nu}\left(D_{\mu}\Phi\right)^*\left(D_{\nu}\Phi\right)
  \label{Complex_rad act}
\ee
where
\be
  D_{\mu}=\partial_{\mu}+iqA_{\mu}
  \label{covariant}
\ee
is the covariant derivative, $q$ is the charge of the field and $A_{\mu}$ is the vector potential.
The complex scalar field is again decomposed into a complete set of basis functions denoted by $\{v_k(r)\}$
\be
  \Phi=\sum_k\left(a_k(t)v_k(r)+ib_k(t)v_k(r)\right).
  \label{Complex_mode}
\ee

Again, since the metric inside and outside the shell have different forms, we split the action into two parts
\be
  S=S_{in}+S_{out}
\ee
where $S_{in}$ and $S_{out}$ are given in Eq.(\ref{Sin}) and Eq.(\ref{Sout}), respectively, with the replacement of the covariant derivative in $S_{out}$. However, from Ref.\cite{Israel}, we can write the vector potential for the collapsing shell as
\be
  (A_{\mu})=\left(\frac{Q}{r},0,0,0\right),
\ee
hence there is no radial component for the vector potential. Thus we can write $D_r\rightarrow\partial_r$ in Eq.(\ref{ApproxAct}) and we can expand Eq.(\ref{Sout}) as
\begin{align}
  S_{out}=&2\pi\int dt\int_{R(t)}^{\infty}dr{}r^2\Big{[}-\frac{1}{1-2GM/r+Q^2/r^2}\times\nonumber\\
     &\left(|\partial_t\Phi|^2+i\frac{qQ}{r}\left[(\partial_t\Phi)\Phi^*-(\partial_t\Phi^*)\Phi\right]-\frac{q^2Q^2}{r^2}|\Phi|^2\right)\nonumber\\
     &+\left(1-\frac{2GM}{r}+\frac{Q^2}{r^2}\right)|D_r\Phi|^2\Big{]}.
\end{align}
Therefore the kinetic term in $S_{in}$ diverges as $(R-R_H)^{-1}$ in this limit and dominates over the softer logarithmically divergent contribution to the kinetic term from $S_{out}$. Similarly, the potential term in $S_{in}$ vanishes in this limit and is sub-dominant compared to the contribution coming from $S_{out}$. Also, if we assume that the charge of the induced radiation is small, we can then ignore the term in $S_{out}$ that is proportional to $(qQ)^2$. Hence,
\begin{align}
  S\sim2\pi\int dt\Big{[}&-\frac{1}{f}\int_0^{R_+}dr{}r^2|\partial_t\Phi|^2\nonumber\\
       &+iqQ\int_{R_+}^{\infty}drr\left[(\partial_t\Phi)\Phi^*-(\partial_t\Phi^*)\Phi\right]\nonumber\\
       &+\int_{R_+}^{\infty}dr{}r^2\left(1-\frac{2GM}{r}+\frac{Q^2}{r^2}\right)|\partial_r\Phi|^2\Big{]}
       \label{ApproxAct}
\end{align}
where we have changed the limits of integration to $R_+$ since we are working in the regime $R(t)\sim R_+$. 

Here the mode coefficients satisfy the commutation relations
\be
  \left[\dot{a}_k(t),a_{k'}(t)\right]=\left[\dot{b}_k(t),b_{k'}(t)\right]=\delta_{k,k'}
\ee
with all others being zero. To ensure this, we take the time derivative of the complex mode expansion, Eq.(\ref{Complex_mode}) to be
\be
  \partial_t\Phi=-i\sum_k\left(\dot{a}_k(t)v_k(r)-i\dot{b}_k(t)v_k(r)\right).
  \label{Complex_mode_t}
\ee

Now, we use the expansion in modes in Eq.(\ref{Complex_mode}) and corresponding derivative in Eq.(\ref{Complex_mode_t}) to write Eq.(\ref{ApproxAct}) as
\begin{align}
  S=\int dt\Big{[}-&\frac{1}{2f}\left[\dot{a}_k{\bf A}_{kk'}\dot{a}_{k'}+\dot{b}_k{\bf A}_{kk'}\dot{b}_{k'}\right]\nonumber\\
     &+qQ\left[\dot{a}_k{\bf B}_{kk'}a_{k'}-\dot{b}_k{\bf B}_{kk'}b_{k'}\right]\nonumber\\
     &+\frac{1}{2}\left[a_k{\bf C}_{kk'}a_{k'}+b_k{\bf C}_{kk'}b_{k'}\right]\Big{]}
\end{align}
where the matrices ${\bf A}$, ${\bf B}$, and ${\bf C}$ are matrices independent of $R(t)$ and are given by
\begin{align}
  {\bf A}_{kk'}&=4\pi\int_0^{R_+}drr^2v_kv_{k'}\\
  {\bf B}_{kk'}&=4\pi\int_0^{R_+}drrv_kv_{k'}\\
  {\bf C}_{kk'}&=4\pi\int_{R_+}^{\infty}drr^2\left(1-\frac{2GM}{r}+\frac{Q^2}{r^2}\right)v'_kv'_{k'}.
\end{align}

To investigate the action of the mode coefficients, we consider the Noether current. The corresponding Noether current is given by
\begin{align}
  N_{\Phi}=&2\pi i\Big{[}\left(-\frac{1}{f}\int_0^{R_H} drr^2(\partial_t\Phi)-iqQ\int_{R_H}^{\infty}drr\Phi\right)\Phi^*\nonumber\\
        &-\left(-\frac{1}{f}\int_0^{R_H} drr^2(\partial_t\Phi)^*+iqQ\int_{R_H}^{\infty}drr\Phi\right)\Phi\Big{]}\nonumber\\
           =&2\pi i\left[\Pi^*\Phi^*-\Pi\Phi\right].
           \label{NC_field}
\end{align}
Using the mode expansion in Eq.(\ref{Complex_mode}) and corresponding derivative in Eq.(\ref{Complex_mode_t}) we can write Eq.(\ref{NC_field}) as
\be
  N_{\Phi}=\frac{1}{f}{\bf A}_{kk'}\left(\dot{a}_{k}a_{k'}-\dot{b}_kb_{k'}\right)+qQ{\bf B}_{kk'}\left(a_ka_{k'}+b_kb_{k'}\right)
  \label{NC}
\ee
Here we see that the the Noether current satisfies the commutation relations
\be
  \left[N_{\Phi},a\right]=a, \hspace{1mm} \text{and} \hspace{1mm} \left[N_{\Phi},b\right]=-b.
\ee
Hence we see that the $a$ modes create particles with $+q$ worth of charge, while the $b$ modes create particles with $-q$ worth of charge. 

Using the standard quantization condition, the wavefunction $\psi(a_k,b_k,t)$ satisfies
\begin{align}
  i\frac{\partial\psi}{\partial t}=&\Big{[}\frac{f}{2}\left[\Pi_k({\bf A}^{-1})_{kk'}\Pi_{k'}+\pi_k({\bf A}^{-1})_{kk'}\pi_{k'}\right]\nonumber\\
     &+\frac{qQ}{2}\left[\Pi_k{\bf B}_{kk''}({\bf A}^{-1})_{k''k'}a_{k'}-\pi_k{\bf B}_{kk''}({\bf A}^{-1})_{k''k'}b_{k'}\right]\nonumber\\
    &+\frac{1}{2}\Big{[}a_k\left(f\frac{q^2Q^2}{2}{\bf B}_{kk''}^2({\bf A}^{-1})_{k''k'}+{\bf C}_{kk'}\right)a_{k'}\nonumber\\
    &+b_k\left(f\frac{q^2Q^2}{2}{\bf B}_{kk''}^2({\bf A}^{-1})_{k''k'}+{\bf C}_{kk'}\right)b_{k'}\Big{]}\Big{]}\psi
\end{align}
where
\be
  \Pi_k=-i\frac{\partial}{\partial a_k}, \hspace{2mm} \text{and} \hspace{2mm} \pi_k=-i\frac{\partial}{\partial b_k}
  \label{conj_mom}
\ee
are the momentum operators conjugate to $a_k$ and $b_k$, respectively. 

To solve this we will assume that the wavefunction can be separated as,
\be
  \psi(c,d,t)=\psi(c,t)\psi(d,t).
\ee
Separating the Schr\"odinger equation then gives the two equations
\begin{align}
  i\frac{\partial\psi(c,t)}{\partial t}=&\Big{[}\frac{f}{2}\Pi_k({\bf A}^{-1})_{kk'}\Pi_{k'}+i\frac{qQ}{2}{\bf B}_{kk''}({\bf A}^{-1})_{k''k'}\delta_{kk'}\nonumber\\
    &+\frac{1}{2}a_k\left(f\frac{q^2Q^2}{2}{\bf B}_{kk''}^2({\bf A}^{-1})_{k''k'}+{\bf C}_{kk'}\right)a_{k'}\Big{]}\psi(c,t)
\end{align}
and
\begin{align}
  i\frac{\partial\psi(d,t)}{\partial t}=&\Big{[}\frac{f}{2}\pi_k({\bf A}^{-1})_{kk'}\pi_{k'}-i\frac{qQ}{2}{\bf B}_{kk''}({\bf A}^{-1})_{k''k'}\delta_{kk'}\nonumber\\
    &+\frac{1}{2}b_k\left(f\frac{q^2Q^2}{2}{\bf B}_{kk''}^2({\bf A}^{-1})_{k''k'}+{\bf C}_{kk'}\right)b_{k'}\Big{]}\psi(c,t)
\end{align}
where we used Eq.(\ref{conj_mom}) in the second term of each equation. Since the two equations are only different by a negative sign in the linear term, we can then again use the principle axis theorem to simultaneously diagonalize the matrices ${\bf A}$, ${\bf B}$ and ${\bf C}$. This then leads to the expressions, written in standard form
\begin{align}
  i\frac{\partial\psi(c,\eta)}{\partial\eta}=&\Big{[}-\frac{1}{2m}\frac{\partial^2}{\partial c^2}+i\frac{y}{m}qQ\nonumber\\
    &+\frac{1}{2}\left(\frac{q^2Q^2}{2}\frac{y^2}{m}+\frac{K}{f}\right)c^2\Big{]}\psi(c,\eta)\nonumber\\
     \approx&\left[-\frac{1}{2m}\frac{\partial^2}{\partial c^2}+i\frac{y}{m}qQ+\frac{1}{2}\frac{K}{f}c^2\right]\psi(c,\eta)
     \label{Harmonic_c}
\end{align}
and
\begin{align}
  i\frac{\partial\psi(d,\eta)}{\partial\eta}=&\Big{[}-\frac{1}{2m}\frac{\partial^2}{\partial d^2}-i\frac{y}{m}qQ\nonumber\\
    &+\frac{1}{2}\left(\frac{q^2Q^2}{2}\frac{y^2}{m}+\frac{K}{f}\right)d^2\Big{]}\psi(d,\eta)\nonumber\\
     \approx&\left[-\frac{1}{2m}\frac{\partial^2}{\partial d^2}-i\frac{y}{m}qQ+\frac{1}{2}\frac{K}{f}d^2\right]\psi(d,\eta)
     \label{Harmonic_d}
\end{align}
where we used the fact that $q$ is small and $f\rightarrow 0$ as $R\rightarrow R_+$, and where $c$ and $d$ are the eigenmodes of $a(t)$ and $b(t)$, and $m$, $y$ and $K$ are the eigenvalues of the matrices ${\bf A}$, ${\bf B}$ and ${\bf C}$, respectively.

To solve Eqs.(\ref{Harmonic_c}) and (\ref{Harmonic_d}), we use the ansatz for the $c$ modes
\be
  \psi(c,t)=e^{\int d\eta qQy/m}\sigma(c,t)=e^{qQy\eta/m}\Sigma(c,t)
  \label{c_Ped}
\ee
similarly the ansatz for the $d$-modes
\be
  \psi(d,t)=e^{-\int d\eta qQy/m}\sigma(d,t)=e^{-qQy\eta/m}\Sigma(d,t)
  \label{d_Ped}
\ee
where we used the fact that $q$, $Q$, $y$ and $m$ are $\eta$ independent. Therefore we can write Eqs.(\ref{Harmonic_c}) and (\ref{Harmonic_d}) as the usual harmonic oscillator equation given in by
\be
  i\frac{\partial\sigma(c,\eta)}{\partial\eta}=\left[-\frac{1}{2m}\frac{\partial^2}{\partial c^2}+\frac{m}{2}\omega^2(\eta)c^2\right]\Sigma(c,\eta)
\ee
and
\be
  i\frac{\partial\sigma(d,\eta)}{\partial\eta}=\left[-\frac{1}{2m}\frac{\partial^2}{\partial d^2}+\frac{m}{2}\omega^2(\eta)d^2\right]\Sigma(d,\eta)
\ee
where $\eta$ and $\omega(\eta)$ are given by Eq.(\ref{eta}) and Eq.(\ref{omega}), respectively. 

The spacetime is static at early times and the initial vacuum state for the modes is the simple harmonic oscillator ground state,
\be
  \Sigma(c(d),\eta=0)=\left(\frac{m\omega_0}{\pi}\right)^{1/4}e^{-m\omega_0c^2(d^2)/2}.
\ee
Then the exact solution at later times is, see Ref.\cite{Pedrosa},
\be
  \Sigma(c,\eta)=e^{i\alpha(\eta)}\left(\frac{m}{\pi\rho^2}\right)^{1/4}\exp\left[i\frac{m}{2}\left(\frac{\rho_{\eta}}{\rho}+\frac{i}{\rho^2}\right)c^2\right]
\ee
where $\rho$ is given by Eq.(\ref{rho}) with initial conditions given in Eq.(\ref{rhoIC}) and the phase $\alpha$ is defined in Eq.(\ref{alpha}), and a similar expression holds for the $d$ modes. 

Note, since the total wave function is given by $\psi(c,d,t)=\psi(c,t)\psi(d,t)$, there is no effect due to the charge for the total wavefunction. Therefore the total occupation number is independent of the charge, hence the induced radiation reduces to the uncharged case. However, the occupation of each mode is dependent on the charge, due to the presence of the exponential term.

Analogous to the uncharged case, consider an observer with detectors that are designed to register particles of different frequencies for the free fields $\phi$ and $\phi^*$ at early times. Such an observer will interpret the wavefunction of a given mode $c$ and $d$ at late times in terms of simple harmonic oscillator states, $\{\varphi_n\}$, at the final frequency $\bar{\omega}$. The number of quanta in eigenmode $c$ and $d$ can be evaluated by decomposing the wavefunctions in Eqs.(\ref{c_Ped}) and (\ref{d_Ped}) in terms of the states, $\{\varphi_n\}$, and by evaluating the occupation number of that mode. To implement this evaluation, we start by writing the wavefunction for a given mode at time $t>t_f$ in terms of the simple harmonic oscillator basis at $t=0$. This is given by
\be
  \psi(c,d,t)=\sum_{n,m}c_n(t)c_m(t)\varphi_n(c)\varphi_m(d)
\ee
where
\be
  c_n=\int dc\varphi_n^*(c)\psi(c,t) \hspace{2mm} \text{and}\hspace{2mm} c_m=\int dd\varphi_n^*(d)\psi(d,t)
  \label{c_n}
\ee
which are the overlap of a Gaussian with the simple harmonic oscillator basis functions. The occupation number at eigenfrequency $\bar{\omega}$ (i.e. in the eigenmode $c$ and $d$) by the time $t>t_f$, are given by the expectation values
\be
  N_c(t,\bar{\omega})=\sum_nn\left|c_n\right|^2 \hspace{2mm} \text{and} \hspace{2mm} N_d(t,\bar{\omega})=\sum_mm\left|c_m\right|^2.
\ee
The occupation number in the eigenmode $c$ is then given by (see Appendix \ref{ch:OccNumCS})
\be
  N_c(t,\bar{\omega})=e^{2qQy\eta/m}\frac{\bar{\omega}\rho^2}{\sqrt{2}}\left[\left(1-\frac{1}{\bar{\omega}\rho^2}\right)^2+\left(\frac{\rho_{\eta}}{\bar{\omega}\rho}\right)^2\right]
  \label{c_occupNum}
\ee
and the occupation number in eigenmode $d$ is then given by
\be
  N_d(t,\bar{\omega})=e^{-2qQy\eta/m}\frac{\bar{\omega}\rho^2}{\sqrt{2}}\left[\left(1-\frac{1}{\bar{\omega}\rho^2}\right)^2+\left(\frac{\rho_{\eta}}{\bar{\omega}\rho}\right)^2\right]
  \label{d_occupNum}
\ee

To analyze the complex scalar field case, we wish to investigate two different scenarios. The first being the non-extremal case. In the non-extremal case there are two horizons, however, as far as the asymptotic observer is concerned the important horizon is the outside horizon. The second is the extremal case. In this case, the two horizons collapse to form one horizon and the domain wall can be characterized by its charge alone. 

\subsubsection{Non-Extremal Case}

Here we will analyze the induced radiation in the non-extremal case. As discussed in the previous section, here we will consider only the situation when $(GM)^2>Q^2$.

\begin{figure}[htbp]
\includegraphics{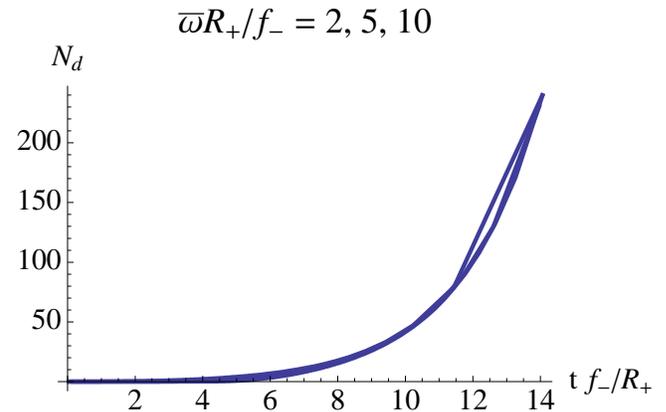}
\caption{Here we plot $N$ versus $tf_-/R_{+}$ for various fixed values of $\bar{\omega}R_{+}/f_-$ for the non-extremal case, with $Q^2=0.4^2(GM)^2$ for the $d$ modes. The curves are lower for higher $\bar{\omega}R_{+}/f_-$.}
\label{Nvst_neg}
\end{figure}

\begin{figure}[htbp]
\includegraphics{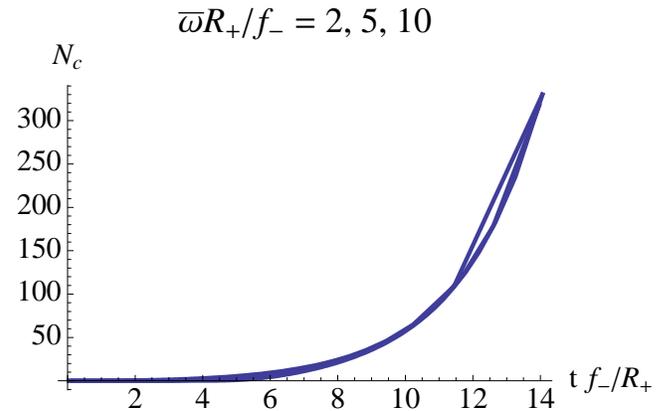}
\caption{Here we plot $N$ versus $tf_-/R_{+}$ for various fixed values of $\bar{\omega}R_{+}/f_-$ for the non-extremal case, with $Q^2=0.4^2(GM)^2$ for the $c$ modes. The curves are lower for higher $\bar{\omega}R_{+}/f_-$.}
\label{Nvst_pos}
\end{figure}

In Figures \ref{Nvst_neg} and \ref{Nvst_pos} we plot $N$ versus $tf_-/R_{+}$ for various fixed values of $\bar{\omega}R_{+}/f_-$. Here we chose that $Q^2=0.4^2(GM)^2$ and $q=0.2(GM)$ for illustration purposes. We see that the occupation number for each mode at any frequency diverges in the infinite time limit when backreaction is not taken into account. 

\begin{figure}[htbp]
\includegraphics{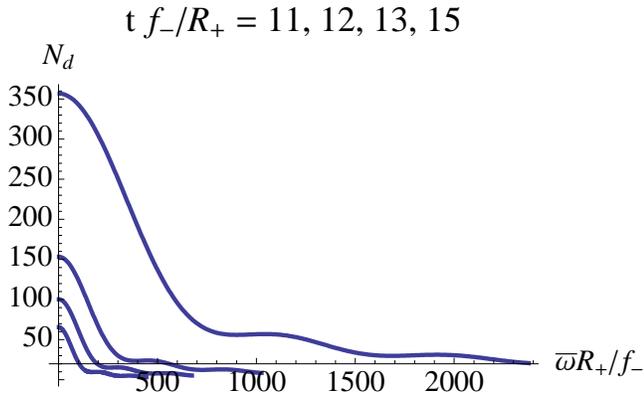}
\caption{Here we plot $N$ versus $\bar{\omega}R_{+}/f_-$ for various fixed values of $tf_-/R_{+}$ for the non-extremal case, with $Q^2=0.4^2(GM)^2$ for the $d$ modes. The occupation number at any frequency grows as $tf_-/R_{+}$ increases.}
\label{Nvsw_neg}
\end{figure}

\begin{figure}[htbp]
\includegraphics{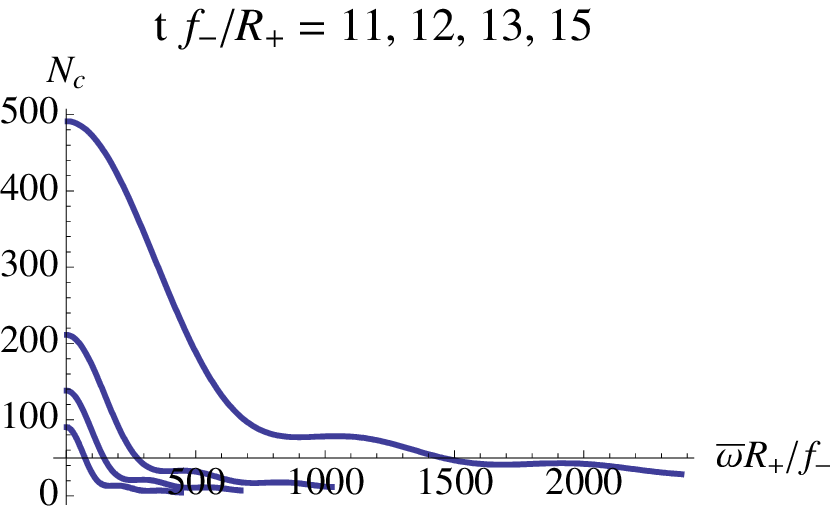}
\caption{Here we plot $N$ versus $\bar{\omega}R_{+}/f_-$ for various fixed values of $tf_-/R_{+}$ for the non-extremal case, with $Q^2=0.4^2(GM)^2$ for the $c$ modes. The occupation number at any frequency grows as $tf_-/R_{+}$ increases.}
\label{Nvsw_pos}
\end{figure}

We have also numerically evaluated the spectrum of mode occupation numbers at any finite time and show the results in Figures \ref{Nvsw_neg} and \ref{Nvsw_pos} for several different values of $tf_-/R_{+}$. We compare the curves in Figures \ref{Nvsw_neg} and \ref{Nvsw_pos} with the occupation numbers for the Planck distribution in Eq.(\ref{N_P}), where again $\beta$ is the inverse temperature. It is clear that the spectrum of occupation numbers is non-thermal. In particular there is no singularity in $N$ at $\omega=0$ at finite time, there are oscillations in $N$. As $tf_-/R_+\rightarrow\infty$, the peak at $\omega=0$ does diverge and the distribution becomes more and more thermal.

\begin{figure}[htbp]
\includegraphics{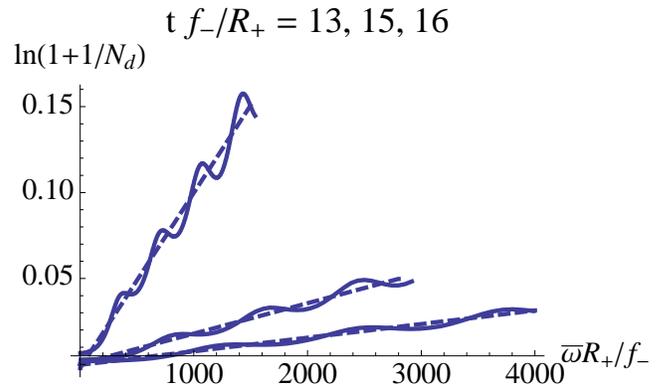}
\caption{Here we plot $\ln(1+1/N)$ versus $\bar{\omega}R_{+}$ for $tf_-/R_+=16$ for the non-extremal case, with $Q^2=0.4^2(GM)^2$, for the $d$ modes. The dashed line shows $\ln(1+1/N_P)$ versus $\bar{\omega}R_{+}/f_-$ where $N_P$ is a Planck distribution. The slope gives $\beta^{-1}$ and the temperature in Eq.(\ref{temp}). The slope of the best fit line decreases as $tf_-/R_+$ increases.}
\label{LnNvsw_neg}
\end{figure}

\begin{figure}[htbp]
\includegraphics{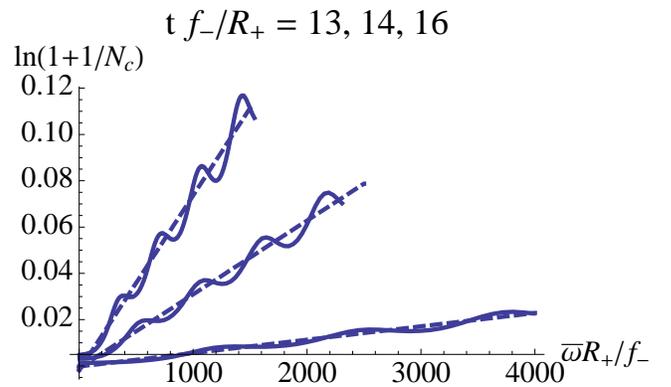}
\caption{Here we plot $\ln(1+1/N)$ versus $\bar{\omega}R_{+}/f_-$ for $tf_-/R_+=16$ for the non-extremal case, with $Q^2=0.4^2(GM)^2$, for the $c$ modes. The dashed line shows $\ln(1+1/N_P)$ versus $\bar{\omega}R_{+}/f_-$ where $N_P$ is a Planck distribution. The slope gives $\beta^{-1}$ and the temperature in Eq.(\ref{temp}). The slope of the best fit line decreases as $tf_-/R_+$ increases.}
\label{LnNvsw_pos}
\end{figure}

We can fit the thermal spectrum to the spectrum of Figures \ref{Nvsw_neg} and \ref{Nvsw_pos}, as shown in Figures \ref{LnNvsw_neg} and \ref{LnNvsw_pos} to obtain the temperatures of radiation
\be
  T\approx0.21\frac{f_-}{R_+}=0.21\frac{\sqrt{(GM)^2-Q^2}}{R_+^2}=1.32 T_H
\ee
for the $c$ modes, and 
\be
  T\approx0.15\frac{f_-}{R_+}=0.15\frac{\sqrt{(GM)^2-Q^2}}{R_+^2}=0.94T_H
\ee
for the $d$ modes, where 
\be
  T_H=\frac{1}{2\pi}\frac{\sqrt{(GM)^2-Q^2}}{R_+^2}
  \label{HawkTemp}
\ee
where we used Eq.(\ref{temp}). We can see that the inverse temperature in each case decreases as $tf_-/R_+$ increases, hence the temperature of the radiation increases over time.

Comparing Figures \ref{Nvst_pos} and \ref{Nvst_neg} we see that for later times the occupation number is greater for the $c$ modes as opposed to the $d$ modes. The occupation number of the $d$ modes still grows as time increases, however, it is subdominant to the occupation number of the $c$ modes. This is expected, since we are assuming the domain wall has a positive charge. Hence the positive charges ($c$ modes) are repelled by the domain wall causing a greater occupation number for late times.

Similarly, comparing Figures \ref{Nvsw_pos} and \ref{Nvsw_neg} we see that as time increased, the occupation number  per frequency is greater for the $c$ modes. Again, the occupation per frequency for the $d$ modes increases as time increases, but it is subdominant to that of the $c$ modes. Again this is expected due to the Coulomb repulsion. 

\subsubsection{Extremal Case}

Here we will analyze the induced radiation in the extremal case. In the extremal case, Eq.(\ref{f}) becomes
\be
  f=\left(1-\frac{Q}{R}\right)^2.
\ee
It was shown in Ref.\cite{WangGreenStoj} that the time dependence of $f$ in the extremal limit is given by $f\approx(Q^2/(t+Q^2))^2$ at late times. 

\begin{figure}[htbp]
\includegraphics{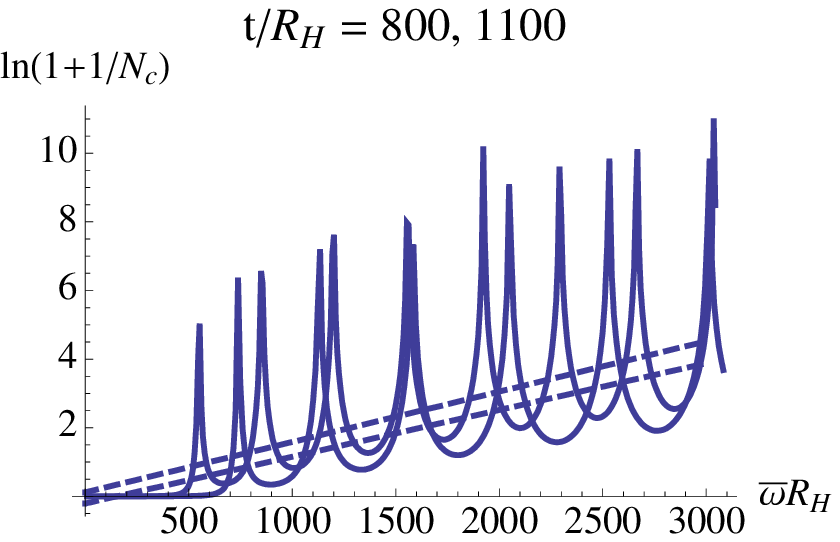}
\caption{Here we plot $\ln(1+1/N)$ versus $\bar{\omega}R_{H}$ for $t/R_H=16$ for the extremal case, with $Q^2=0.4^2(GM)^2$, for the $c$ modes. The dashed line shows $\ln(1+1/N_P)$ versus $\bar{\omega}R_{H}$ where $N_P$ is a Planck distribution. The slope gives $\beta^{-1}$. The slope of the best fit line decreases as $t/R_H$ increases.}
\label{LnNvsw_Expos}
\end{figure}

\begin{figure}[htbp]
\includegraphics{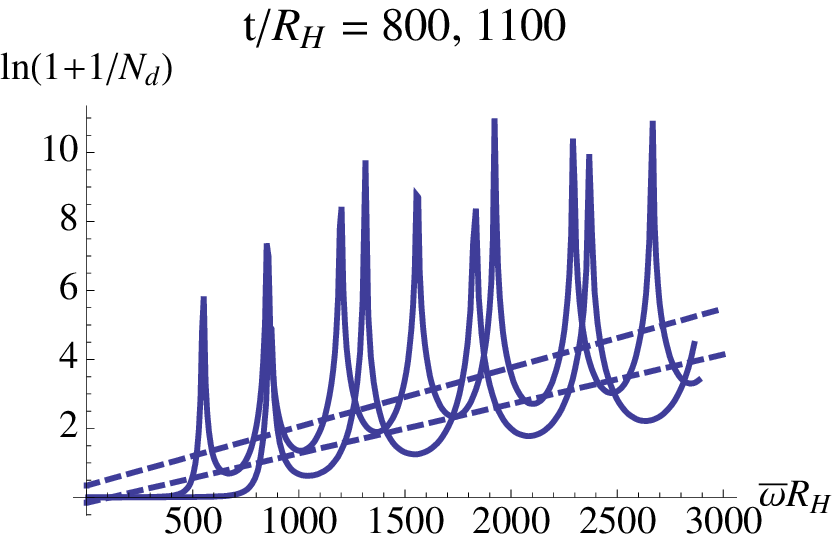}
\caption{Here we plot $\ln(1+1/N)$ versus $\bar{\omega}R_{H}$ for $t/R_H=16$ for the extremal case, with $Q^2=0.4^2(GM)^2$, for the $c$ modes. The dashed line shows $\ln(1+1/N_P)$ versus $\bar{\omega}R_{H}$ where $N_P$ is a Planck distribution. The slope gives $\beta^{-1}$. The slope of the best fit line decreases as $t/R_H$ increases.}
\label{LnNvsw_Exneg}
\end{figure}

In Figures \ref{LnNvsw_Expos} and \ref{LnNvsw_Exneg} we plot $\ln(1+1/N)$ versus $\bar{\omega}R_H$ for various fixed values of $t/R_H$ for $c$ modes and $d$ modes, respectively. Here we compare the inverse temperature versus that in Figure \ref{LnNvsw_pos}. Similarly to the non-extremal case, here we see that the inverse temperature decreases as $t/R_H$ increases. Now from Eq.(\ref{Schrod_eta}), since the time derivative of the wavefunction on the right-hand side is with respect to $\eta$, $\omega$ is the mode frequency with respect to $\eta$ and not with respect to time $t$. Eq.(\ref{eta}) tells us that the frequency in $t$ is $(Q^2/(t+Q^2))^2$ times the frequency in $\eta$, and at time $t_f$, this implies
\be
  \omega^{(t)}=\left(\frac{Q^2}{t+Q^2}\right)^2\bar{\omega}
  \label{Omega_t}
\ee
where the superscript $(t)$ on $\omega$ refers to the fact that this frequency is with respect to $t$. This rescaling of the frequency implies that the temperature for the asymptotic observer (with time coordinate $t$) can be obtained by finding the ``best fit temperature" $\beta^{-1}$ and then rescaling by $(Q^2/(t+Q^2))^2$. So the temperature seen by the asymptotic observer is
\be
  T=\left(\frac{Q^2}{t+Q^2}\right)^2\beta^{-1}(t_f).
  \label{Temp}
\ee
We can fit the thermal spectrum as shown in Figures \ref{LnNvsw_Expos} and \ref{LnNvsw_Exneg} to obtain the temperature of radiation. Here we see that the inverse temperature increases as $t/R_H$ increase, meaning that the temperature decreases over time. Hence the temperature of the extremal black hole decreases as $t/R_H$ increases. Once the shell crosses the horizon, the temperature will go to zero. This is consistent with the fact that the temperature is given by  Eq.(\ref{HawkTemp}).

\section{Conclusion}

Here we consider the induced radiation from a collapsing Reisner-Nordstr\"om domain wall using a semi-classical analysis. To do so we calculate the occupation of the induced radiation, $N(t,\bar{\omega})$. The occupation number strongly depends on time, due to the fact that one measures the time dependent frequency, $\bar{\omega}$. The time dependence of the frequency depends on the observer's time, therefore one has the freedom to chose a particular observer for the analysis. In this paper we have chosen to investigate the induced radiation from the view point of an asymptotic observer, an observer located at infinity. 

First we considered the case when a neutral scalar field is coupled to the background of the collapsing domain wall. In this case there is no qualitative difference from that of the uncharged (Schwarzschild) case. This is expected since uncharged matter does not couple to the charge of the domain wall, hence as far as the scalar field is concerned the background metric is just Schwarzschild. There are, of course, minor quantitative differences between the two cases, since the background metrics are different. However, these differences are not important here, since they do not shed light on the impact of the charge of the domain wall on the induced radiation. Comparing with the results of Ref.\cite{VachStojKrauss}, we see that the results of the scalar field here are consistent with the results found here.

To take the charge into account, we coupled a complex scalar field to the background of the Reisner-Nordstr\"om domain wall. Here we considered two different cases, first the non-extremal case and second the extremal case, respectively. First we will discuss the non-extremal case.

In the non-extremal case, we find that as time increases the occupation number for charges with the same sign as the domain wall (here represented by the $c$ modes) become dominant over that of the charges with sign opposite to that of the domain wall (represented here by the $d$ modes). This is an expected result due to the Coulomb repulsion, the like charges will be repelled by the wall causing a greater flux of these charges seen by the asymptotic observer, while the opposite charges will be attracted to the domain wall causing a smaller flux. By fitting the temperature, we find that the temperature of the induced radiation is on the same order as that of the Hawking radiation for late times. 

In the extremal case, we find that there is radiation induced due to the collapse of the domain wall, due to the time-dependent nature of the metric. Unlike the non-extremal case, the temperature of the radiation decreases as $t/R_H$ increases. This is in agreement with Eq.(\ref{HawkTemp}), where the black hole can obtain absolute zero. 

Here we demonstrated that as time increases, or as the domain wall approaches the horizon, the occupation number of the modes which carry charge with the same sign as to the domain wall become dominant over that of the occupation of the modes which carry charge with opposite sign to that of the domain wall. As discussed, this is expected due to the Coulomb repulsion (attraction) due to the charges. Therefore, when the domain wall gets very close to the horizon, the flux of radiation measured by the asymptotic observer will be completely dominated by radiation with the same sign as that of the domain wall. The radiation distribution function is not quite thermal, though it becomes thermal when the collapsing object reaches its own horizon. We call such radiation Hawking-like or pre-Hawking radiation (as opposed to thermal Hawking radiation from a pre-existing horizon). However, once the horizon is formed the Hawking effect takes over and the radiation is completely thermal.

\section*{Acknowledgments}

The author wishes to thank both Dejan Stojkovic and Ulrich Baur for useful discussions and helpful discussions.

\appendix

\section{$\rho$ Equation}
\label{rho_annal}

In the range $t<0$, $\omega$ is a constant and the solution to Eq. (\ref{rho}) is 
\be
  \rho(\eta)=\frac{1}{\sqrt{\omega_0}}.
\ee

In the range $0<t<t_f$, we do not have an analytic solution but we can derive certain useful properties. First note that in terms of $\eta$
\be
  \omega^2=\frac{\omega_0^2}{f_{-}(1-\eta/R_{+})}.
\ee
Then the equation for $\rho$ after rescaling can be written as:
\be
  \frac{d^2g}{d\eta'^2}=(\omega_0R_{+})^2\left[\frac{g}{f_{-}(1-\eta')}-\frac{1}{g^3}\right]
  \label{g_eq}
\ee
where $\eta'=\eta/R_{+}$, $g=\sqrt{\omega_0}\rho$. The boundary conditions are
\be
  g(0)=1, \hspace{2mm} \frac{dg(0)}{d\eta'}=0.
\ee
The last term with the $1/g^3$ becomes singular as $g\rightarrow0$. Let us consider another equation with the $1/g^3$ replaced by something better behaved. For example
\be
  \frac{d^2h}{d\eta'^2}=(\omega_0R_{+})^2\left[\frac{h}{f_{-}(1-\eta')}-h\right]
  \label{h_eq}
\ee
with boundary conditions
\be
  h(0)=1, \hspace{2mm} \frac{dh(0)}{d\eta'}=0.
  \label{h_IC}
\ee
Eq. (\ref{h_eq}) implies that $h(\eta')$ is monotonically decreasing as long as $h(\eta')>0$. Furthermore, it is decreasing faster than the solution for $g$ as long as $g<1$, since the $1/g^3$ term in Eq. (\ref{g_eq}) is a larger ``repulsive" force than the $h$ term in Eq. (\ref{h_eq}). So 
\be
  h(\eta')\geq g(\eta')
\ee
for all $\eta'$ such that $g(\eta')>0$. 

Eq. (\ref{h_eq}) with initial conditions (\ref{h_IC}) can be solved in terms of degenerate hypergeometric functions. For us, the important point is that the solution for $h$ is positive for all $\eta'$ and, in particular, $h(1)>0$ for all the values of $\omega_0R_{+}$ we have checked. Therefore $g(\eta')$ is positive, at least for a wide range of $\omega_0R_{+}$. 

Let $g_1=g(1)\not=0$. Then the equation for $g$ can be expanded near $\eta'=1$.
\be
  \frac{d^2g}{d\eta'^2}\sim-(\omega_0R_{+})^2\left[\frac{g_1}{f_{-}(1-\eta')}-\frac{1}{g_1^3}\right].
\ee
This shows that
\be
  \frac{dg}{d\eta'}\sim(\omega_0R_{+})^2\frac{g_1}{f_{-}}\ln(1-\eta')\rightarrow-\infty
\ee
as $\eta'\rightarrow1$. 

Hence $\rho(\eta=R_{+})$ is strictly positive and finite while $\rho_{\eta}(\eta=R_{+})=-\infty$ for finite and non-zero $\omega_0$. Since $g=\sqrt{\omega_0}\rho$, and $g\rightarrow1$ for $\omega_0\rightarrow0$, we also see that $\rho\rightarrow0$ and $\rho_{\eta}\rightarrow0$ as $\omega_0\rightarrow0$.

In the range $t_f<t$, $\omega$ is a constant. However, the solution for $\rho$ is not a constant, unlike in the range $t<0$, since the constant solution $1/\sqrt{\omega(t_f)}$ does not necessarily match up with $\rho(t_f-)$ to ensure a continuous solution. Yet it is easy to check that in this region $\dot{N}=0$ and so there is no change in the occupation numbers. So we need only find $N(t_f-,\bar{\omega})$ to determine $N(t\rightarrow\infty,\bar{\omega})$. 

\section{Number of Particles Radiated as a Function of Time, for Scalar Field}
\label{ch:OccNumS}

We use the simple harmonic oscillator basis states but at a frequency $\bar{\omega}$ to keep track of the different $\omega$'s in the calculation. To evaluate the occupation numbers at time $t>t_f$, we need only set $\bar{\omega}=\omega(t_f)$. So 
\be
  \phi_n(b)=\left(\frac{m\bar{\omega}}{\pi}\right)^{1/4}\frac{e^{-m\bar{\omega}b^2/2}}{\sqrt{2^nn!}}{\cal{H}}_n(\sqrt{m\bar{\omega}}b)
\ee
where ${\cal{H}}_n$ are Hermite polynomials. Then Eq. (\ref{c_n}) together with Eq. (\ref{PedWave}) gives
\begin{align}
  c_n&=\left(\frac{1}{\pi^2\bar{\omega}\rho^2}\right)^{1/4}\frac{e^{i\alpha}}{\sqrt{2^nn!}}\int d\xi e^{-P\chi^2/2}{\cal{H}}_n(\xi)\nonumber\\
  &\equiv\left(\frac{1}{\pi^2\bar{\omega}\rho^2}\right)^{1/4}\frac{e^{i\alpha}}{\sqrt{2^nn!}}I_n
\end{align}
where
\be
  P=1-\frac{i}{\bar{\omega}}\left(\frac{\rho_{\eta}}{\rho}+\frac{i}{\rho^2}\right).
\ee

To find $I_n$ consider the corresponding integral over the generating function for the Hermite polynomials
\begin{align}
  J(z)&=\int d\xi e^{-P\xi^2/2}e^{-z^2+2z\xi}\nonumber\\
    &=\sqrt{\frac{2\pi}{P}}e^{-z^2(1-2/P)}.
\end{align}
Since
\begin{align}
  e^{-z^2+2z\xi}&=\sum_{n=0}^{\infty}\frac{z^n}{n!}{\cal{H}}(\xi)\\
  \int d\xi e^{-P\xi^2/2}{\cal{H}}_n(\xi)&=\frac{d^n}{dz^n}J(z)\Big{|}_{z=0}.
\end{align}
Therefore
\be
  I_n=\sqrt{\frac{2\pi}{P}}\left(1-\frac{2}{P}\right)^{n/2}{\cal{H}}_n(0).
\ee
Since 
\be
  {\cal{H}}_n(0)=(-1)^{n/2}\sqrt{2^nn!}\frac{(n-1)!!}{\sqrt{n!}}, \hspace{2mm} \text{n=even}
\ee
and ${\cal{H}}_n(0)=0$ for odd $n$, we find the coefficients $c_n$ for even values of $n$,
\be
  c_n=\frac{(-1)^{n/2}e^{i\alpha}}{(\bar{\omega}\rho^2)^{1/4}}\sqrt{\frac{2}{P}}\left(1-\frac{2}{P}\right)^{n/2}\frac{(n-1)!!}{\sqrt{n!}}.
\ee
For odd $n$, $c_n=0$.

Next we find the number of particles produced. Let 
\be
  \chi=\left|1-\frac{2}{P}\right|.
\ee
Then
\begin{align}
  N(t,\bar{\omega})&=\sum_{n=\text{even}}n\left|c_n\right|^2\nonumber\\
      &=\frac{2}{\sqrt{\bar{\omega}\rho^2}|P|}\chi\frac{d}{d\chi}\sum_{n=\text{even}}\frac{(n=1)!!}{n!!}\chi^n\nonumber\\
      &=\frac{2}{\sqrt{\bar{\omega}\rho^2}|P|}\chi\frac{d}{d\chi}\frac{1}{\sqrt{1-\chi^2}}\nonumber\\
      &=\frac{2}{\sqrt{\bar{\omega}\rho^2}|P|}\frac{\chi^2}{(1-\chi^2)^{3/2}}.
\end{align}
Inserting the expressions for $\chi$ and $P$, leads to 
\be
  N(t,\bar{\omega})=\frac{\bar{\omega}\rho^2}{\sqrt{2}}\left[\left(1-\frac{1}{\bar{\omega}\rho^2}\right)^2+\left(\frac{\rho_{\eta}}{\bar{\omega}\rho}\right)^2\right].
\ee

\section{Number of Particles Radiated as a Function of Time, for Complex Scalar Field}
\label{ch:OccNumCS}

We use the simple harmonic oscillator basis states but at a frequency $\bar{\omega}$ to keep track of the different $\omega$'s in the calculation. To evaluate the occupation numbers at time $t>t_f$, we need only set $\bar{\omega}=\omega(t_f)$. So 
\be
  \phi_n(c)=\left(\frac{m\bar{\omega}}{\pi}\right)^{1/4}\frac{e^{-m\bar{\omega}c^2/2}}{\sqrt{2^nn!}}{\cal{H}}_n(\sqrt{m\bar{\omega}}b)
\ee
where ${\cal{H}}_n$ are Hermite polynomials. Then Eq. (\ref{c_n}) together with Eq. (\ref{PedWave}) gives
\begin{align}
  c_n&=\left(\frac{1}{\pi^2\bar{\omega}\rho^2}\right)^{1/4}\frac{e^{i\alpha}}{\sqrt{2^nn!}}e^{qQy\eta/m}\int d\xi e^{-P\chi^2/2}{\cal{H}}_n(\xi)\nonumber\\
  &\equiv\left(\frac{1}{\pi^2\bar{\omega}\rho^2}\right)^{1/4}\frac{e^{i\alpha}}{\sqrt{2^nn!}}e^{qQy\eta/m}I_n
\end{align}
where
\be
  P=1-\frac{i}{\bar{\omega}}\left(\frac{\rho_{\eta}}{\rho}+\frac{i}{\rho^2}\right).
\ee

To find $I_n$ consider the corresponding integral over the generating function for the Hermite polynomials
\begin{align}
  J(z)&=\int d\xi e^{-P\xi^2/2}e^{-z^2+2z\xi}\nonumber\\
    &=\sqrt{\frac{2\pi}{P}}e^{-z^2(1-2/P)}.
\end{align}
Since
\begin{align}
  e^{-z^2+2z\xi}&=\sum_{n=0}^{\infty}\frac{z^n}{n!}{\cal{H}}(\xi)\\
  \int d\xi e^{-P\xi^2/2}{\cal{H}}_n(\xi)&=\frac{d^n}{dz^n}J(z)\Big{|}_{z=0}.
\end{align}
Therefore
\be
  I_n=\sqrt{\frac{2\pi}{P}}\left(1-\frac{2}{P}\right)^{n/2}{\cal{H}}_n(0).
\ee
Since 
\be
  {\cal{H}}_n(0)=(-1)^{n/2}\sqrt{2^nn!}\frac{(n-1)!!}{\sqrt{n!}}, \hspace{2mm} \text{n=even}
\ee
and ${\cal{H}}_n(0)=0$ for odd $n$, we find the coefficients $c_n$ for even values of $n$,
\be
  c_n=\frac{(-1)^{n/2}e^{i\alpha}}{(\bar{\omega}\rho^2)^{1/4}}\sqrt{\frac{2}{P}}\left(1-\frac{2}{P}\right)^{n/2}e^{qQy\eta/m}\frac{(n-1)!!}{\sqrt{n!}}.
\ee
For odd $n$, $c_n=0$.

Next we find the number of particles produced. Let 
\be
  \chi=\left|1-\frac{2}{P}\right|.
\ee
Then
\begin{align}
  N_c(t,\bar{\omega})&=e^{2qQy\eta/m}\sum_{n=\text{even}}n\left|c_n\right|^2\nonumber\\
      &=e^{2qQy\eta/m}\frac{2}{\sqrt{\bar{\omega}\rho^2}|P|}\chi\frac{d}{d\chi}\sum_{n=\text{even}}\frac{(n=1)!!}{n!!}\chi^n\nonumber\\
      &=e^{2qQy\eta/m}\frac{2}{\sqrt{\bar{\omega}\rho^2}|P|}\chi\frac{d}{d\chi}\frac{1}{\sqrt{1-\chi^2}}\nonumber\\
      &=e^{2qQy\eta/m}\frac{2}{\sqrt{\bar{\omega}\rho^2}|P|}\frac{\chi^2}{(1-\chi^2)^{3/2}}.
\end{align}
Inserting the expressions for $\chi$ and $P$, leads to 
\be
  N_c(t,\bar{\omega})=e^{2qQy\eta/m}\frac{\bar{\omega}\rho^2}{\sqrt{2}}\left[\left(1-\frac{1}{\bar{\omega}\rho^2}\right)^2+\left(\frac{\rho_{\eta}}{\bar{\omega}\rho}\right)^2\right].
\ee

Similarly, following the same steps for the $d$-modes leads to
\be
  N_d(t,\bar{\omega})=e^{-2qQy\eta/m}\frac{\bar{\omega}\rho^2}{\sqrt{2}}\left[\left(1-\frac{1}{\bar{\omega}\rho^2}\right)^2+\left(\frac{\rho_{\eta}}{\bar{\omega}\rho}\right)^2\right].
\ee

\end{document}